\def\beq{\begin{equation}}
\def\eeq{\end{equation}}
\def\blg{\begin{align}}
\def\elg{\end{align}}
\def\beg{\begin{gather}}
\def\eeg{\end{gather}}
\def\bea{\begin{eqnarray}}
\def\eea{\end{eqnarray}}
\def\bed{\begin{displaymath}}
\def\eed{\end{displaymath}}
\def\bef{\begin{figure} \begin{center}}
\def\eef{\end{center} \end{figure}}

\def\no{\noindent}
\def\nn{\nonumber}

\def\cao{\c c\~ao\ }

\def\1{\'{\i}}

\documentclass[preprint,pre,aps]{revtex4-1}
\usepackage{amssymb,amsmath,graphics,graphicx}

\begin{document}

\title{\bf Bursting synchronization in networks with long-range coupling mediated by a diffusing chemical substance}

\author{R. L. Viana $^1$ \footnote{Corresponding author. e-mail: viana@fisica.ufpr.br}, A. M. Batista $^2$, C. A. S. Batista $^{2}$, J. C. A. de Pontes $^3$, F. A. dos S. Silva $^1$, and S. R. Lopes $^1$}

\affiliation{1. Departamento de F\'{\i}sica, Universidade Federal do Paran\'a, 81531-990, Curitiba, Paran\'a, Brazil. \\ 2. Departamento de Matem\'{a}tica e Estat\'{\i}stica, Universidade Estadual de Ponta Grossa, 84032-900, Ponta Grossa, Paran\'{a}, Brazil. \\ 3. Coordena\c c\~ao de Automa\c c\~ao Industrial, Universidade Tecnol\'ogica Federal do Paran\'a, 84016-210, Ponta Grossa, Paran\'a, Brazil.}

\date{\today}   

\begin{abstract}
Many networks of physical and biological interest are characterized by a long-range coupling mediated by a chemical which diffuses through a medium in which oscillators are embedded. We considered a one-dimensional model for this effect for which the diffusion is fast enough so as to be implemented through a coupling whose intensity decays exponentially with the lattice distance. In particular, we analyzed the bursting synchronization of neurons described by two timescales (spiking and bursting activity), and coupled through such a long-range interaction network. One of the advantages of the model is that one can pass from a local (Laplacian) type of coupling to a global (all-to-all) one by varying a single parameter in the interaction term. We characterized bursting synchronization using an order parameter which undergoes a transition as the coupling parameters are changed through a critical value. We also investigated the role of an external time-periodic signal on the bursting synchronization properties of the network. We show potential applications in the control of pathological rhythms in biological neural networks.
\end{abstract}

\maketitle

\section{Introduction}

In many problems of physical and biological interest we consider nonlinear oscillators whose interaction is mediated by a substance which is secreted by the cells and diffuses along the inter-cellular medium, being absorbed by the cells. Moreover, the rate of secretion depends on the cell dynamics, as well as the rate of absorption. In this way the dynamics of the cells are effectively coupled by the diffusing substance, leading to a non-local coupling type which depends on the details of the diffusion process as well as the dynamics of the oscillators themselves, leading to a complex system which displays a wealth of dynamical behaviors, like periodic and chaotic regimes, bifurcations, crises, destruction of tori, among other features.

An outstanding biological example of interaction mediated by a chemical is the collective behavior of brain cells responsible by the circadian rhythm. The circadian rhythm is a daily periodicity (roughly a $24$ hr cycle) of physiological, biochemical, and behavioral processes in living beings \cite{winfree}. It is produced in mammals by specialized cells (circadian clocks) belonging to the suprachiasmatic nucleus (SCN) of the anterior hypothalamus \cite{dunlap}. The SCN consists of multiple, single-cell circadian clocks which, when synchronized, produce a coherent circadian output that regulate overt rhythms \cite{vdp,miller,hastings}. 

The circadian master clock of the SCN is entrained by the daily light-dark cycle, which acts via retina-to-SCN neural pathways \cite{liu}. Hence, to obtain a coordinated circadian rhythm, the master clock cell must be coupled to the other cells in the SCN so as to synchronize them to its own rhythm. The chemical coupling between circadian clocks can be described by means of a chemical ($\gamma$-aminobutyric acid, or GABA, for short) which is both secreted and absorbed by clock cells immersed in some the intra-cellular medium. The coupling, in this case, is non-local in the sense that it takes into account cells which are not necessarily close to each other. An extreme situation belonging to this general category is the so-called global coupling, in which each cell interacts with the average concentration of the chemical in the inter-cellular medium due to all the other cells \cite{liu1}. 

Another situation in which this kind of coupling is potentially important is chemotaxis, which is the influence of chemicals on the motion of somatic cells, bacteria, and other single or multiple-cell organisms \cite{chemobook}. These chemicals diffuse in the environment of the cells and the corresponding concentration gradients are responsible for the movement of the cells. For example, bacteria find glucose by swimming towards regions of higher concentration of such food molecules in the environment \cite{livro2}. Alternatively, they can also flee from poison (such as phenol) according to their local concentration. Moreover, sperm can move towards the egg during fertilization, thanks to concentration gradients of aminoacids, in a ligand/receptor interaction. 

The chemotactic ability of slime molds like {\it Dictyostelium} is responsible for starving amoebae to gather to form multicellular bodies \cite{livro3}. During most of their lives, these slime molds are individual unicellular protists living in similar habitats and feeding on microorganisms. In the absence of food, however, they release signal molecules (DIF-1, short for Differentiation Inducing Factor) into their environment, so that they can find other amoeba and create swarms. In other words, when a chemical signal is secreted, they assemble into a cluster which acts as a single organism \cite{slim}.

A model for nonlocal coupling mediated by a diffusing chemical was proposed by Kuramoto, in which the equations governing the time evolution of the oscillators can be coupled by using the concentration of a substance which diffuses through the medium in which the oscillators are embedded \cite{kuramoto}. If the chemical diffuses in a timescale much faster than the oscillator period the coupling, although involves virtually all oscillators like in the global case, depends on the distance between oscillators in an exponentially decaying way \cite{batto,nakao}. This approach has been used in studies of cell interaction \cite{batto1} and neural oscillators \cite{sakaguchi}. In the latter example the chemical secreted and absorbed by the neurons is a neurotransmitter which mediates the coupling among neurons. This coupling, on its way, depends also on the dynamics of the individual neurons, since it determines the propagation of electrical impulses along neuronal circuits in the brain. 

This neuronal activity (i.e., the evolution of the action potential) presents a fast timescale characterized by repetitive spiking and a slow timescale with bursting activity, where neuron activity alternates between a quiescent state and spiking trains \cite{belykh05}. There are many models for this spiking-bursting behavior, comprising differential equations like the Hindmarsch-Rose model \cite{rose} and discrete-time maps, as the Rulkov map \cite{rulkov01,rulkov03}. 

An assembly of coupled bursting neurons exhibit many self-organized phenomena. We are particularly interested in bursting synchronization, for which the neurons burst at approximately the same time, even though their spiking behavior may be not synchronized itself \cite{belykh05}. Thanks to the slow timescale we may regard each bursting neuron as a phase oscillator, on defining a bursting phase and a corresponding frequency (its time rate) \cite{ivanchenko04}. 

On the other hand, neurophysiologists argue that bursting synchronization plays a key role in some pathologies like Parkinson's disease, essential tremor, and epilepsies \cite{park}. Therefore, the question of how to suppress this synchronization acquires a practical importance in terms of the control of undesirable neuronal rhythms. There have been proposed some alternatives to implement such a control through deep brain stimulation techniques. One of these is the addition of a time-periodic external signal of small amplitude and given frequency \cite{rosenblum,ivanchenko04}. We have considered such a scheme in scale-free networks \cite{pontes3} and regular lattices with a power-law coupling \cite{pontes2}. Another deep brain stimulation technique was proposed by Rosenblum and Pikowsky, and consists of a time-delayed feedback signal applied to specific cortex areas \cite{rosenblum,tukhlina}. We have recently studied this technique in scale-free networks \cite{nosso}. 

In this work we propose to study the control of bursting synchronization in a neuronal network with a long-range coupling mediated by a diffusing substance, using the Kuramoto model in the adiabatic limit of a rapidly diffusing chemical \cite{kuramoto,nakao,batto}. We shall use a time-periodic external signal and a time-delayed feedback signal so as to suppress bursting synchronization whenever it occurs, and study this phenomenon from the point of view of our coupling model.

This paper is organized as follows: in Section 2 we outline a model for long-range coupling intermediated by a diffusing substance. In Section 3 we particularize the model to one-dimensional oscillator chains and coupled map lattices. In Section 4 we consider the bursting dynamics as described by the Rulkov map, and a long-range coupled network of Rulkov neurons. In Section 5 we investigate the bursting synchronization of coupled Rulkov neurons according to the nonlocal coupling model. Section 6 deals with the control of synchronized bursting rhythms by application of an external time-periodic driving and a delayed feedback signal. Our conclusions are left to the last Section.

\section{Long-range coupling mediated by a diffusing substance}

Chemical coupling mediated by a diffusing substance can be mathematically described by a model proposed by Kuramoto leading to nonlocal couplings \cite{kuramoto}. In this model the state variables of each oscillator influence the secretion of a chemical substance obeying a diffusion equation \cite{batto,nakao,batto1,sakaguchi}. The rate of absorption depends on the local concentration of this substance at each cell position. In the following we will deal with two classes of vectors, which are represented with a different notation: (i) positions ${\vec r}$ in a $d$-dimensional Euclidean space, to which the oscillators belong; (ii) state variables ${\mathbf X} = {(x_1, x_2, \ldots x_M)}^T$ in a $M$-dimensional phase space of the dynamical variables characterizing the state of the system at a given time $t$.

\begin{figure}
\begin{center}
\includegraphics[width=0.8\textwidth,clip]{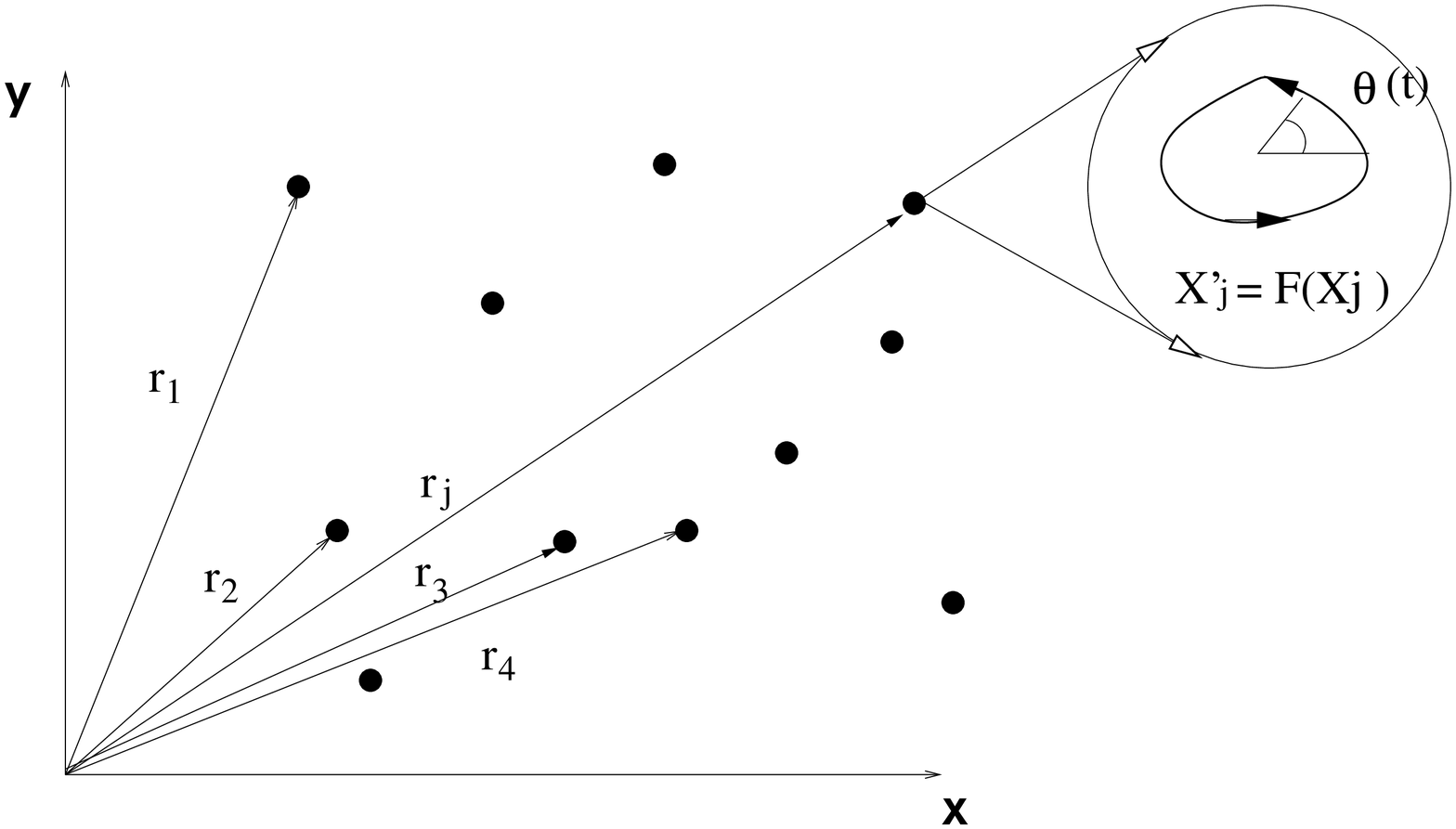}
\caption{\label{chemfig} Schematic figure of spatially distributed phase oscillators.}
\end{center}
\end{figure}

There are $N$ oscillator cells located at discrete positions ${\vec r}_j$, where $j = 1, 2, \cdots N$, in the $d$-dimensional Euclidean space; and ${\mathbf X}_j$ is the state variable for each oscillator, whose time evolution is governed by the vector field ${\mathbf F}({\mathbf X}_j)$ [Fig. \ref{chemfig}]. The oscillators are not supposed to be identical, though, for they can have slightly different parameters. 

We suppose that the time evolution is affected by the local concentration of a chemical, denoted as $A({\vec r},t)$, through a time-dependent coupling function ${\mathbf g}$:
\begin{equation}
\label{oscil}
\frac{d{\mathbf X}_j}{dt} = {\mathbf F}({\mathbf X}_j) + {\mathbf g}(A({\vec r},t)),
\end{equation}
\noindent whereas the chemical concentration satisfies a diffusion equation of the form
\begin{equation}
\label{chem}
\varepsilon \frac{\partial A({\vec r})}{\partial t} = - \eta A({\vec r},t) + D \nabla^2 A({\vec r},t) + \sum_{k=1}^{N} h({\mathbf X}_j) \delta({\vec r} - {\vec r}_k),
\end{equation}
\noindent where $\varepsilon \ll 1$ is a small parameter representing the fact that diffusion occurs in a timescale faster than the intrinsic period of individual oscillators; $\eta$ is a phenomenological damping parameter, and $D$ is a diffusion coefficient. The diffusion equation above has a source term $h$ which depends on the oscillator state at the discrete spatial positions ${\vec r}_j$: this means that each oscillator secrets the chemical with a rate depending on the current value of its own state variable. 

According to Ref. \cite{kuramoto} we assume that the diffusion is very fast, compared with the oscillator period, such that we may set $\varepsilon{\dot A} = 0$. This makes the concentration of the mediating chemical to relax immediately to a stationary value that can be written in the following form:
\begin{equation}
\label{green}
A({\vec r}_j) = \sum_{k=1}^{N} \sigma({\vec r}_j - {\vec r}_k) h({\mathbf X}_k),
\end{equation}
\noindent where $\sigma({\vec r}_j - {\vec r}_k)$ is a Green function (or an interaction kernel), which is the solution of
\begin{equation}
\label{kernel}
(\eta - D \nabla^2)\sigma({\vec r}_j - {\vec r}) = \delta({\vec r}_j).
\end{equation}

Since we have eliminated adiabatically the concentration of the diffusing chemical, on substituting (\ref{green}) into Eq. (\ref{chem}) we can obtain an equation expressing the nonlocal coupling in the adiabatic approximation
\begin{equation}
\label{oscil1}
\frac{d{\mathbf X}_j}{dt} = {\mathbf F}({\mathbf X}_j) + {\mathbf g}\left(\sum_{k=1}^{N} \sigma({\vec r}_j - {\vec r}_k) h({\mathbf X}_k) \right).
\end{equation}

If ${\mathbf g}$ is a linear function of ${\mathbf X}_j$ (but not necessarily of the positions ${\vec r}_j$) we can write
\begin{equation}
\label{oscila1}
\frac{d{\mathbf X}_j}{dt} = {\mathbf F}({\mathbf X}_j) + \sum_{k=1}^{N} \sigma({\vec r}_j - {\vec r}_k) {\mathbf g}(h({\mathbf X}_k)).
\end{equation}
\no in such a way that we distinguish among some cases of interest: (i) the {\it linear coupling}, for which
\beq
\label{linear}
{\mathbf g}(h({\mathbf X}_k)) = {\mathbf A}{\mathbf X}_k,
\eeq
\no where ${\mathbf A}$ is a $M \times M$ matrix indicating which variables of the oscillators are coupled; (ii) the {\it future coupling}, where
\beq
\label{future}
{\mathbf g}(h({\mathbf X}_k)) = {\mathbf A}{\mathbf F}({\mathbf X}_k),
\eeq
\no and (iii) the {\it nonlinear coupling}, such that
\beq
\label{nonlinear}
{\mathbf g}(h({\mathbf X}_k)) = {\mathbf A}{\mathbf H}({\mathbf X}_k),
\eeq
\no where ${\mathbf H}$ is a nonlinear function of its arguments.

The equation for linear coupling, for instance, is thus
\begin{equation}
\label{oscillinear}
\frac{d{\mathbf X}_j}{dt} = {\mathbf F}({\mathbf X}_j) + \sum_{k=1}^{N} \sigma({\vec r}_j - {\vec r}_k) {\mathbf A}{\mathbf X}_k ,
\end{equation}
\no As an example, let us consider each cell as undergoing a time evolution governed by the Hindmarch-Rose equations \cite{rose}, which also describe neurons with spiking and bursting activity. For this system we have  $M=3$ and 
\beq
\label{hr}
{\mathbf X} = 
\left( \begin{array}{c}
x \\
y \\
z 
\end{array} \right) ,
\qquad
{\mathbf F} = 
\left( \begin{array}{c}
y + ax^2 - x^3 - z + I \\
1 - bx^2 - y \\
r[s(x-\chi) - z]
\end{array} \right),
\eeq
\no where $a$, $b$, $I$, $r$, $s$, and $\chi$ are model parameters.

Coupling these equations through the $x$-variable amounts to choose
\beq
\label{A}
{\mathbf A} = 
\left( \begin{array}{ccc}
\varepsilon & 0 & 0 \\
0 & 0 & 0 \\
0 & 0 & 0
\end{array} \right),
\eeq
\no where $\varepsilon$ is the coupling strength; such that the equation (\ref{oscillinear}) for the linear coupling reads, in this case
\bea
\nn
{\dot x}_j & = & y_j + a x_j^2 - x_j^3 - z_j + I + \varepsilon \sum_{k=1}^{N} \sigma({\vec r}_j - {\vec r}_k) x_k, \\
\label{hmlinear}
{\dot y}_j & = & 1 - b x_j^2 - y_j , \qquad (j = 1, 2, \ldots N) \\
\nn
{\dot z}_j & = & r[s(x_j - \chi) - z_j] .
\eea
\no If we were to couple these equations through the $y$-variable, we would have to use a different matrix, namely
\beq
\label{A3}
{\mathbf A} = 
\left( \begin{array}{ccc}
0 & 0 & 0 \\
0 & \varepsilon & 0 \\
0 & 0 & 0
\end{array} \right),
\eeq
\no and so on. The corresponding equation for future coupling is obtained simply by changing ${\mathbf X}_j$ by ${\mathbf F}({\mathbf X}_j)$ in (\ref{oscillinear}).

As an example of nonlinear coupling, let us consider the case of $M = 1$, for which ${\mathbf X}_j$ is a single phase $\theta_j \in [0, 2\pi)$, the vector function ${\mathbf F}({\mathbf X}_j)$ being the corresponding frequency $\omega_j$ (different for each oscillator, in general). In this case ${\mathbf A}$ reduces to a scalar coupling strength $\varepsilon$ and we can choose as the nonlinear coupling function 
\beq
{\mathbf H}({\mathbf X}_k) = \sin(\theta_j - \theta_k),
\eeq
\no yielding a nonlinearly coupled Kuramoto model \cite{kuramotobook,rogers}
\beq
\label{kuramodel}
{\dot\theta}_j = \omega_j + \varepsilon \sum_{k=1}^{N} \sigma({\vec r}_j - {\vec r}_k) \sin(\theta_j - \theta_k).
\eeq

From the Fourier transform of Eq. (\ref{kernel}), the interaction kernel can be written as
\begin{equation}
\label{oscil0}
\sigma({\mathbf r}_j - {\mathbf r}) = \frac{1}{{(2\pi)}^d} \int d^d{\mathbf q} \frac{e^{i{\mathbf q}\cdot({\mathbf r}_j - {\mathbf r})}}{\eta + D {\vert{\mathbf q}\vert}^2}.
\end{equation}
\noindent If the system is isotropic, the kernel becomes a function of the distance $R \equiv \vert{\mathbf r}_j - {\mathbf r}\vert$ only, and can be expressed as
\begin{equation}
\label{sigma123}
\sigma(R) =
\begin{cases}
C e^{-\gamma R}, & \text{if $d = 1$,} \\
C K_0(\gamma R), & \text{if $d = 2$,} \\
C \frac{e^{-\gamma R}}{\gamma R}, & \text{if $d = 3$}
\end{cases}
\end{equation}
\noindent where $K_0$ is the modified Bessel function of the second kind and order $0$, the constant $\gamma$ is the inverse of the coupling length and is given by
\begin{equation}
\label{gama}
\gamma = \sqrt{\frac{\eta}{D}},
\end{equation}
\noindent and the constant $C$ is determined from the normalization condition
\begin{equation}
\label{norm}
\int d^d{\mathbf r} \sigma({\mathbf r}) = 1.
\end{equation}

\section{One-dimensional lattices with long-range coupling}

The most elementary application of the model of nonlocal coupling mediated by a diffusing chemical is a one-dimensional regular lattice of $N$ (an odd number) fixed and equally spaced oscillators with non-local interactions given by the kernel (\ref{sigma123}) for $d = 1$. Assuming that the distance between consecutive lattice sites is a constant $\Delta$, and supposing periodic boundary conditions 
\begin{equation}
\label{boundary}
{\mathbf X}_j = {\mathbf X}_{j \pm N'}, \qquad N' = \frac{N-1}{2},
\end{equation}
\noindent we can write $|{\vec r}_j - {\vec r}_k| = (j-k)\Delta \equiv \ell \Delta$ by changing the summation index from $k = 1, 2, \ldots N$ to $\ell = j-k$, such that $\ell = \pm 1, \pm 2, \ldots \pm N'$. This leaves us with $2N'+1 = N$ sites, each of them with a distance $\ell$ from any site $j$. The Green function is thus
\beq
\label{sigma2}
\sigma(|{\mathbf r}_j - {\mathbf r}|) = C e^{-\gamma \Delta \ell}.
\eeq

On excluding self-interactions, or the coupling of any site with itself, we replace the sum over $k$ in Eq. (\ref{oscillinear}) with two sums, one over $\ell = 1, 2, \ldots N'$ and other over $\ell = -1, -2, \ldots -N'$. In the latter sum we can change index again $m = -\ell$ and then replace $m$ by $\ell$ since they are dummy indexes. Hence the first sum considers $k = j - \ell$ and the second one $k = j + \ell$, such that we can group them together into a single summation, giving for the linear coupling the following expression
\begin{equation}
\label{oscillinear1}
\frac{d{\mathbf X}_j}{dt} = {\mathbf F}({\mathbf X}_j) + C \sum_{\ell=1}^{N'} e^{-\gamma \Delta \ell} {\mathbf A}\left( {\mathbf X}_{j-\ell} + {\mathbf X}_{j+\ell} \right).
\end{equation}

The normalization constant is determined from (\ref{norm}) which, in this case, becomes a summation rather than an integral
\beq
\sum_{k=1}^{N} \sigma(|{\mathbf r}_j - {\mathbf r}|) = 1.
\eeq
\no Making the same changes of index as in the previous paragraph we obtain
\beq
\label{C}
C = {\left\lbrack 2 \sum_{\ell=1}^{N'} e^{-\gamma \Delta \ell} \right\rbrack}^{-1}. 
\eeq

On returning to our previous example of linearly coupled Hindmarch-Rose equations the $x$-coupled system is then
\bea
\nn
{\dot x}_j & = & y_j + a x_j^2 - x_j^3 - z_j + I + \varepsilon C \sum_{\ell=1}^{N'} e^{-\gamma\Delta\ell} (x_{j-\ell} + x_{j+\ell}), \\
\label{hmlinear1}
{\dot y}_j & = & 1 - b x_j^2 - y_j, \qquad (j = 1, 2, \ldots N) \\
\nn
{\dot z}_j & = & r[s(x_j - \chi) - z_j] .
\eea
\no whereas, in the one-dimensional case, the nonlinearly coupled Kuramoto model (\ref{kuramodel}) becomes
\beq
\label{kuramodel1}
{\dot\theta}_j = \omega_j + \varepsilon C \sum_{\ell=1}^{N'} e^{-\gamma\Delta\ell} \left\lbrack \sin(\theta_j - \theta_{j-\ell}) - \sin(\theta_j - \theta_{j+\ell}) \right\rbrack.
\eeq

It is instructive to explore the limiting cases of the nonlocal coupling. If $\gamma$ goes to zero then,
\begin{equation}
\label{normlim1}
C = \frac{1}{2N'} = \frac{1}{N-1},
\end{equation}
\noindent and we have a global type of coupling 
\begin{equation}
\label{oscil3}
\frac{d{\mathbf X}_j}{dt} = {\mathbf F}({\mathbf X}_j) + {\mathbf A} {\overline{\mathbf X}},
\end{equation}
\noindent where each oscillator interacts with the mean field of all sites (except itself)
\begin{equation}
\label{mean}
{\overline{\mathbf X}} = \frac{1}{N-1} \sum_{k=1, k\ne j}^{N} {\mathbf X}_{k} .
\end{equation}

In the limit of $\gamma$ large, the exponentials in the Green function decay very fast with the lattice distance $\ell$, such that only the term with $\ell = 1$ contributes significantly to the summations. This gives for the normalization constant
\beq
\label{Cinf}
C \approx \frac{1}{2 e^{-\gamma\Delta}}
\eeq
\no and the coupling term takes into account effectively only the nearest neighbors of a given site
\beq
\label{oscil4}
\frac{d{\mathbf X}_j}{dt} = {\mathbf F}({\mathbf X}_j) + \frac{1}{2} {\mathbf A} \left( {\mathbf X}_{j-1} + {\mathbf X}_{j+1} \right) 
\eeq
\no which is the usual local (or laplacian) coupling. 

\section{Nonlocally coupled Rulkov networks}

In this and the forthcoming sections we present, as an application of nonlocally couplings mediated by a diffusing substance, a model of coupled bursting neurons. Depending on the phenomena we are interested to investigate, the mathematical description of biological neurons may require models ranging from dozens of complicated differential equations to simple integrate-and-fire one-dimensional models \cite{izhi}. In the following we are going to study collective phenomena involving the spiking and bursting activity of neurons, for which there is a slow bursting modulating the fast action-potential spiking. 

\subsection{Local dynamics}

Since we are interested on finding the conditions for an assembly of neurons to burst synchronously, in particular the coupling parameters necessary to achieve this goal, the fine details of the neuron dynamics are not essential to the network model. Hence, instead of continuous-time models like the Hindmarch-Rose model, which can be slower to simulate computationally, we choose to use discrete-time models instead, which are fast and reliable \cite{dhamala04,miguel}. 

A simple model but which nevertheless presents all the essential behavior of bursting neurons is the Rulkov map \cite{rulkov01,rulkov03}
\begin{eqnarray}
\label{rulkovx}
x_{n+1} & = & \frac{\alpha}{1 + x_n^2} + y_n, \\
\label{rulkovy}
y_{n+1} & = & y_n - \sigma x_n - \beta,
\end{eqnarray}
\noindent where $x_n$ is the fast and $y_n$ is the slow dynamical variable. The first variable has a dynamical behavior emulating the spiking-bursting activity of a neuron, depending on the parameter $\alpha$, whereas the latter variable undergoes a slow evolution because of the small values taken on by the parameters $\sigma$ and $\beta$, which model the action of external dc bias current and synaptic inputs on a given isolated neuron \cite{rulkov03}. 

\begin{figure}
\includegraphics[width=0.7\textwidth,clip]{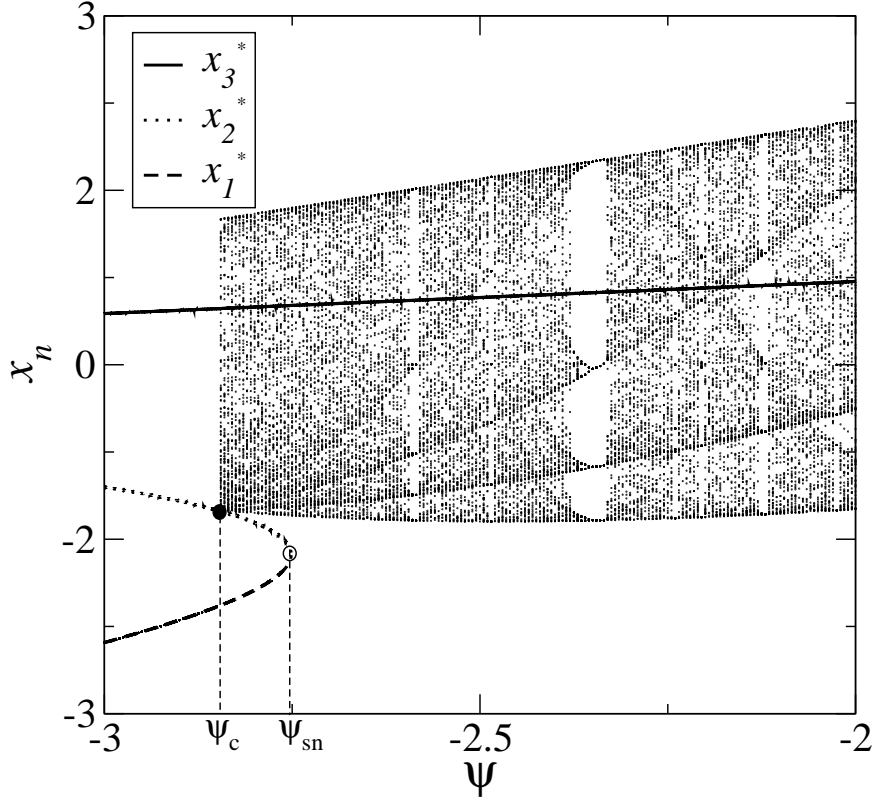}
\caption{\label{adic} Bifurcation diagram for the map (\ref{rulkovapr}) when $\alpha = 4.1$. $x^*_{1,2,3}$ are the fixed points, and we indicated the location of the saddle-node bifurcation ($\psi_{SN}$) and interior crisis ($\psi_C$).}
\end{figure}

\begin{figure}
\includegraphics[width=0.7\textwidth,clip]{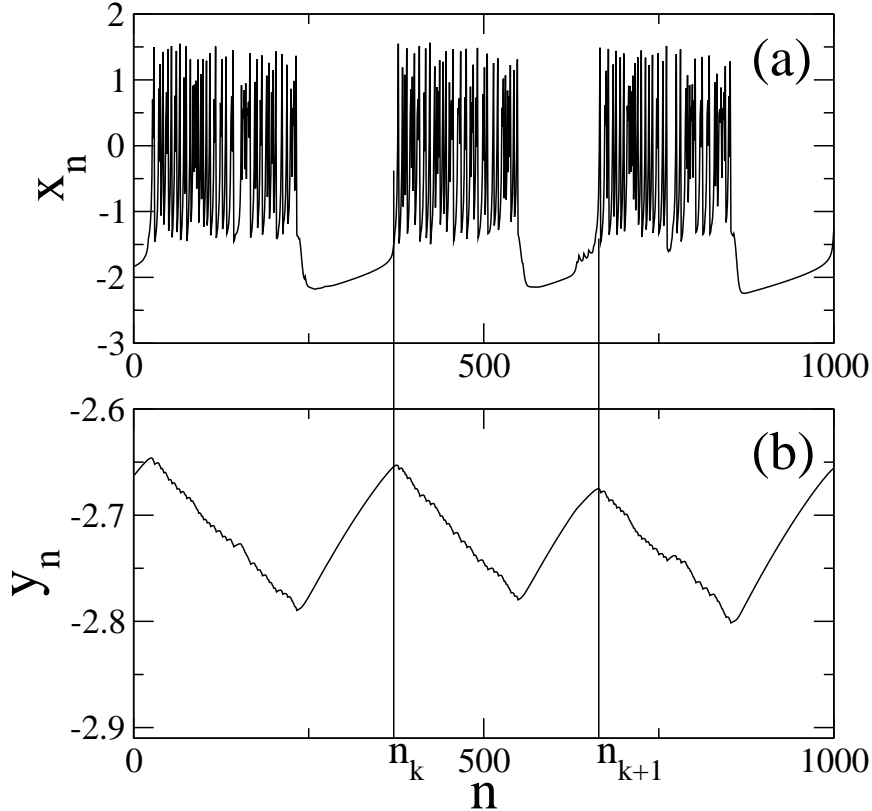}
\caption{\label{buster} Time evolution of the (a) fast and (b) slow variables in the Rulkov map (\ref{rulkovx})-(\ref{rulkovy}) for $\alpha = 4.1$, $\sigma = \beta = 0.001$. We also indicate by $n_k$ the local maxima of the slow variable, used to obtain a phase for the bursting dynamics.}
\end{figure}

We suppose a transient chaotic behavior for the characteristic spiking of the fast variable $x_n$, what is accomplished by choosing the values of the parameter $\alpha$ within the interval $[4.1, 4.3]$ [Fig. \ref{buster}(a)]. The bursting timescale, on the other hand, comes about by the influence of the slow variable $y_n$. This can be understood by using a simple argument: since, from Eq. (\ref{rulkovx}), $y_n$ represents a small input on the fast variable dynamics its effect can be approximated by a constant value $\psi$. The  resulting one-dimensional map, 
\beq
\label{rulkovapr}
x_{n+1} = \frac{\alpha}{1 + x_n^2} + \psi,
\eeq
\no can have either one, two, or three fixed points $x^*_{1,2,3}$, depending on the value of the input $\psi$. 

As the latter approaches a critical value $\psi_{SN}$ the fixed points $x_{1,2}^*$ (one stable and another unstable) undergo a saddle-node bifurcation, such that, for $\psi \gtrsim \psi_{sn}$, however, the fixed points $x^*_{1,2}$ disappear [Fig. \ref{adic}]. For values of $\psi > \psi_{C}$ there is also a chaotic attractor that, provided $\psi_{C} < \psi < \psi_{SN}$, coexists with the stable fixed point attractor. Actually, at $\psi = \psi_{C}$ the chaotic attractor collides with the unstable fixed point $x_1^*$ and is destroyed through a boundary crisis [Fig. \ref{adic}] \cite{newref}. The bursting regime then results from a hysteresis between the stable fixed point (quiescent evolution) and the chaotic oscillations (fast sequence of spikes). 

We consider a given burst to begin when the slow variable $y_n$, which presents nearly regular saw-teeth oscillations, has a local maximum, in well-defined instants of time we call $n_k$ [Fig. \ref{buster}(b)]. We can define a phase describing the time evolution within each burst and varying from $0$ to $2\pi$ as $n$ evolves from $n_k$ to $n_{k+1}$:
\begin{equation}
\label{phase}
\varphi(n) = 2\pi k + 2\pi \frac{n - n_k}{n_{k+1} - n_k}.
\end{equation}

The duration of the chaotic burst, $n_{k+1} - n_k$, depends on the variable $x_n$ and fluctuates in an irregular fashion when $x_n$ undergoes a chaotic evolution. There follows that the bursting phase rate also varies with time, such that we look at the bursting frequency defined by
\begin{equation}
\label{frequency}
\Omega = \frac{\varphi(n) - \varphi(0)}{n}.
\end{equation}

\subsection{Network structure}

When the Euclidean distance between neurons does not play a significant role, the corresponding networks may be treated from the graph-theoretical point of view. However, once we regard those neurons as embedded in a three-dimensional lattice (the brain, where they are connected by axons and dendrites), it is convenient to use a lattice embedded in a Euclidean space \cite{rozenfeld}. 

Such higher-dimensional lattices can be very difficult to work with in terms of computer simulations, specially if long-range interactions are present and the number of neurons is large. However, good insights are expected to come from simpler models, which can nevertheless retain some of the general characteristics of higher-dimensional lattices. It is from this point of view that we use one-dimensional lattices with $N$ neurons, each of them being described by the Rulkov map (\ref{rulkovx}-\ref{rulkovy}).

The main assumption we make is that the neurons are coupled through the release and absorption of a neurotransmitter, which is able to diffuse quickly in the medium in which the neurons are embedded. The release of the neurotransmitter is influenced by the fast dynamics, that is, once the action potential is spiking there are liberated neurotransmitter molecules in the cell environment. These molecules are then absorbed by other neurons and are supposed to affect their fast dynamics, making them to spike in a slightly different way.

This would be certainly a rather crude model to describe information transmission through neuron cells, for it does not incorporate other features of neuron electrical and chemical coupling. However, this model, in spite of its simplicity, suffices to exemplify in which sense does a coupling mediated by a diffusing substance can affect coherent rhythms and other collective phenomena, what is potentially interesting for other applications, like chemotaxis.

The model of nonlocal coupling mediated by a diffusing substance, outlined in the previous Section, can be straightforwardly translated to the language of coupled map lattices, for which both space and time are discrete. We will use the notation ${\mathbf X}^{(j)}_n$ for the state variable vector at the site $j = 1, 2, \ldots N$ (in a one-dimensional lattice with periodic boundary conditions) and time $n = 0, 1, 2, \ldots$. The vector field ${\mathbf F}$ specifies now the map equations, and the interpretation of the remaining variables is the same as for oscillator chains.

The equation for linear coupling of maps becomes
\begin{equation}
\label{cmllinear}
{\mathbf X}_{n+1}^{(j)} = {\mathbf F}({\mathbf X}_{n}^{(j)}) + C \sum_{\ell=1}^{N'} e^{-\gamma \Delta \ell} {\mathbf A}\left( {\mathbf X}_n^{(j-\ell)} + {\mathbf X}_n^{(j+\ell)} \right) ,
\end{equation}
\no the future coupled map lattice being obtained by replacing ${\mathbf X}_n^{(j)}$ by ${\mathbf F}({\mathbf X}_n^{(j)})$ in the coupling term. The normalization constant is still given by Eq. (\ref{C}). 

Let each neuron to undergo a discrete time evolution governed by the Rulkov map (\ref{rulkovx})-(\ref{rulkovy}), for which $M=2$ and 
\beq
\label{rul}
{\mathbf X} = 
\left( \begin{array}{c}
x \\
y 
\end{array} \right) 
\qquad
{\mathbf F} = 
\left( \begin{array}{c}
\frac{\alpha}{1 + x^2} + y \\
y - \sigma x - \beta
\end{array} \right).
\eeq
\no where $\alpha$, $\sigma$ and $\beta$ are model parameters. Since there is always some biological diversity in a neuron assembly it is reasonable to choose slightly different values for the map parameters. 

In order to avoid chance correlations of bursting for uncoupled neurons we choose randomly the $\alpha$ parameter (influencing the fast dynamics) within the interval $[4.1, 4.3]$ with a uniform probability. The slow dynamics parameters $\sigma$ and $\beta$ were all set up to the same (small) values, namely $0.001$, since their possible differences are not likely to affect the results as the parameter $\alpha$ does. 

Since we assume that the release and absorption of neurotransmitters affect the fast dynamics, we couple the Rulkov maps through the $x$-variable. For the sake of simplicity, we use a linear coupling. In some other applications, however, a future coupling would be better so as to keep the coupled variables within prescribed ranges, a feature that is warranted in linear couplings only if the coupling strength is weak enough. These assumptions amount to choose the following connection matrix
\beq
\label{A2}
{\mathbf A} = 
\left( \begin{array}{cc}
\varepsilon & 0 \\
0 & 0 
\end{array} \right),
\eeq
\no where $\varepsilon$ is the coupling strength. The resulting linearly coupled Rulkov map lattice becomes
\bea
\label{rmx}
x_{n+1}^{(j)} & = & \frac{\alpha^{(j)}}{1 + {(x_n^{(j)})}^2} + y_n^{(j)} + 
\varepsilon C \sum_{\ell=1}^{N'} e^{-\gamma\Delta\ell} \left\lbrack x_n^{(j-\ell)} + x_n^{(j+\ell)} \right\rbrack, \\ 
\label{rmy}
y_{n+1}^{(j)} & = & y_n^{(j)} - \sigma x_n^{(j)} - \beta,
\eea
\no where we include a superscript to the $\alpha$ variables to denote the different values they may take throughout the network.

\section{Bursting synchronization of coupled Rulkov neurons}

The coupled Rulkov map lattice (\ref{rmx})-(\ref{rmy}) cannot exhibit a completely synchronized state, 
\[
x_n^{(1)} = x_n^{(2)} = \cdots = x_n^{(N)}, \qquad y_n^{(1)} = y_n^{(2)} = \cdots = y_n^{(N)}.
\]
\no For this to occur it would be necessary that the completely synchronized state be a possible solution of Eqs. (\ref{rmx})-(\ref{rmy}), stable under infinitesimal perturbations along directions transversal to this state. One of the conditions for this is that the individual maps must be identical. This is a condition too stringent to be obeyed by realistic neuron models, for which there must be some degree of diversity in parameters. To mirror this fact, in our numerical simulations the values of the parameter $\alpha$ of each neuron were randomly chosen inside a given interval, provided we always have bursting. However, as a consequence, a completely synchronized state is not possible for our model. 

This does not necessarily mean that the neural network cannot present some degree of coherent behavior. The neuron bursting phases, for example, can synchronize through the interaction provided by the coupling. We measure this effect by computing the mean field.
\begin{equation}
\label{meanfield}
M_n =  \frac{1}{N} \sum_{j=1}^N x_n^{(j)} .
\end{equation}  
\noindent If the neurons are weakly coupled, they burst at different times in a non-coherent fashion, and the mean field fluctuates irregularly with small amplitudes. Oppositely, if the neurons burst synchronously (i.e. they start bursting at approximately the same times) a nonzero mean field is formed and $M_n$ presents regular oscillations of comparatively large amplitude. Only the slow dynamics becomes coherent as the neurons burst synchronously. The fast (spiking) dynamics remains incoherent and do not contribute to the mean field dynamics, which is kept close to a periodic regime \cite{ivanchenko04}.

\begin{figure}
\includegraphics[width=0.8\textwidth,clip]{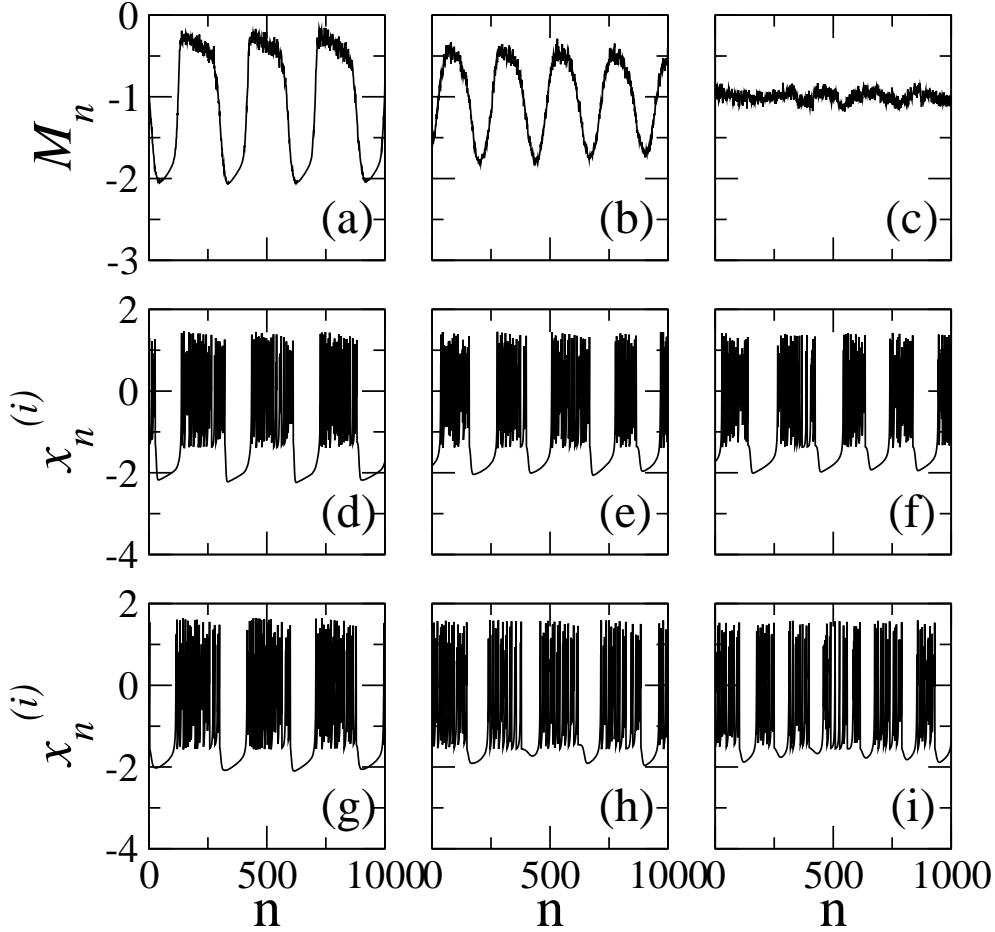}
\caption{\label{campo} Time evolution of the mean field (a) and the fast variables of two selected neurons [(d) and (g)] belonging to a linearly coupled lattice of $N = 251$ Rulkov neurons with $\gamma = 0.005$. (b,e,h) are the corresponding results for $\gamma = 0.0125$, and (c,f,i) for $\gamma = 0.05$. The remaining parameters are $\sigma=\beta=0.001$ and the parameter $\alpha$ is randomly chosen within the interval $[4.1, 4.3]$, with coupling strength $\varepsilon = 0.1$}
\end{figure}

When the neurons are more globally coupled (small $\gamma$) the mean field exhibits large-amplitude oscillations  indicating that the neurons are bursting at approximately the same times [Fig. \ref{campo}(a)]. This can also be shown by inspecting the individual time evolution of the fast variable [Figs. \ref{campo}(d) and (g)] where we select two neurons with different values of $\alpha$, which start bursting at nearly the same time instants. Since such behavior is likely to be observed for all other neurons this corresponds to a coherent output for the entire network.  

On the other hand, when the neurons are more locally coupled, what is achieved by choosing larger values of $\gamma$, only the nearest neighbors contribute appreciably to the coupling, and the mean field $M_n$ displays only small-amplitude fluctuations, as illustrated by [Fig. \ref{campo}(c)]. In this case different neurons burst out of phase [Figs. \ref{campo}(f) and (i)]. An intermediate value of $\gamma$ parameter shows a nearly coherent neuron bursting [Fig. \ref{campo} (b)], for the neurons burst at slightly different times [Figs. \ref{campo}(e) and (h)]. This suggests that, as the coupling becomes increasingly global, there is a transition to bursting synchronization. 

In order to characterize bursting synchronization, however, the mean field is not precise enough, such that we use instead Kuramoto's order parameter \cite{kuramotobook}
\begin{equation}
\label{orderpar}
z_n = R_n \exp\left(i \Phi_n \right) \equiv \frac{1}{N} \sum_{j=1}^{N} \exp(i \varphi^{(j)}_n),
\end{equation}
\noindent where $R_n$ and $\Phi_n$ are the amplitude and angle, respectively, of the order parameter. If the neurons burst at exactly the same times, their phases superimpose coherently such that $R_n = 1$. By way of contrast, if the neurons burst in completely uncorrelated times the corresponding phases $\varphi^{(j)}_n$ would add to a near-zero value. Hence $R_n$ can be used as a numerical diagnostic of bursting synchronization.

\begin{figure}
\includegraphics[width=0.7\textwidth,clip]{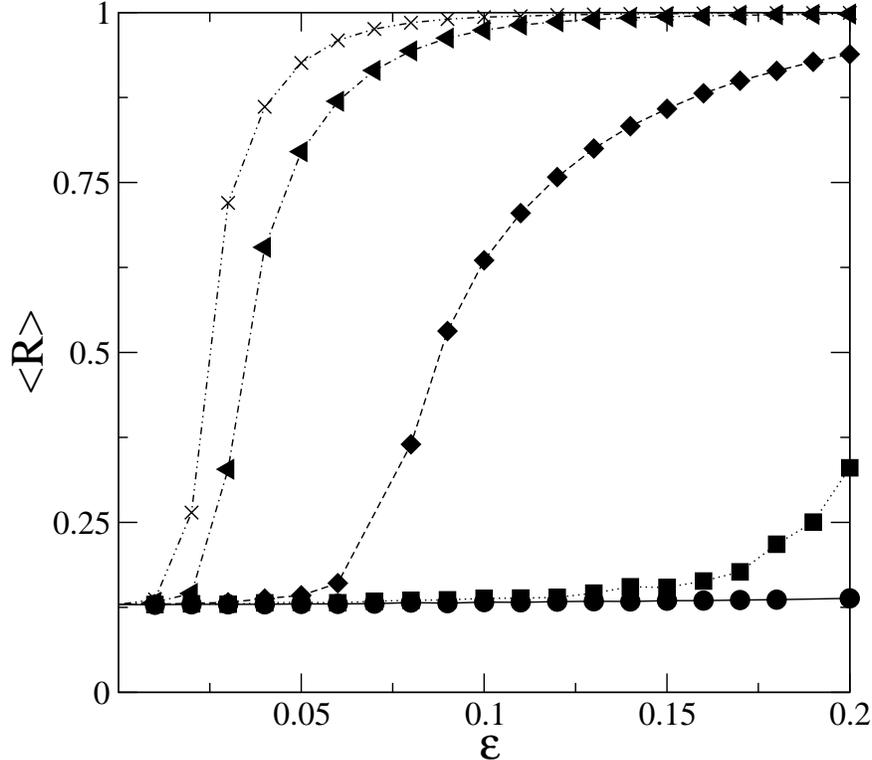}
\caption{\label{order} Time-averaged order parameter as a function of the coupling strength for $\gamma$ equal to $0.025$ (circles), $0.020$ (squares), $0.0125$ (diamonds), $0.005$ (triangles) and $0.002$ (crosses). We remaining parameters are the same as in Fig. \ref{campo}.}
\end{figure}

In Fig. \ref{order} we plot the time-averaged order parameter magnitude $<R>$ (after the transients have died out) as a function of the coupling strength $\varepsilon$, for different values of $\gamma$. For very small values of $\varepsilon$ the order parameters take on very small, yet nonzero, values. Hence the weakly coupled neurons are not likely to produce coherent bursting activity. The order parameter magnitude would not be equal to zero, even for very weak coupling, since the lattice size $N$ is finite. As we consider larger lattices this value is expected to approach zero. 

As a general trend we see that a transition to synchronized bursting as the coupling strength increases. Specially for very small values of $\gamma$, this transition is similar to that exhibited by the well-known the Kuramoto model (or globally coupled oscillators) [see Eq. (\ref{kuramodel})] \cite{kuramotobook}. In fact, the latter would correspond to the $\gamma\rightarrow 0$ limit of our model. Provided it is kept small enough, as $\gamma$ increases, this transition occurs for higher values of the coupling strength. However, for a further increase, this transition no longer seem to occur. 

\begin{figure}
\includegraphics[width=0.7\textwidth,clip]{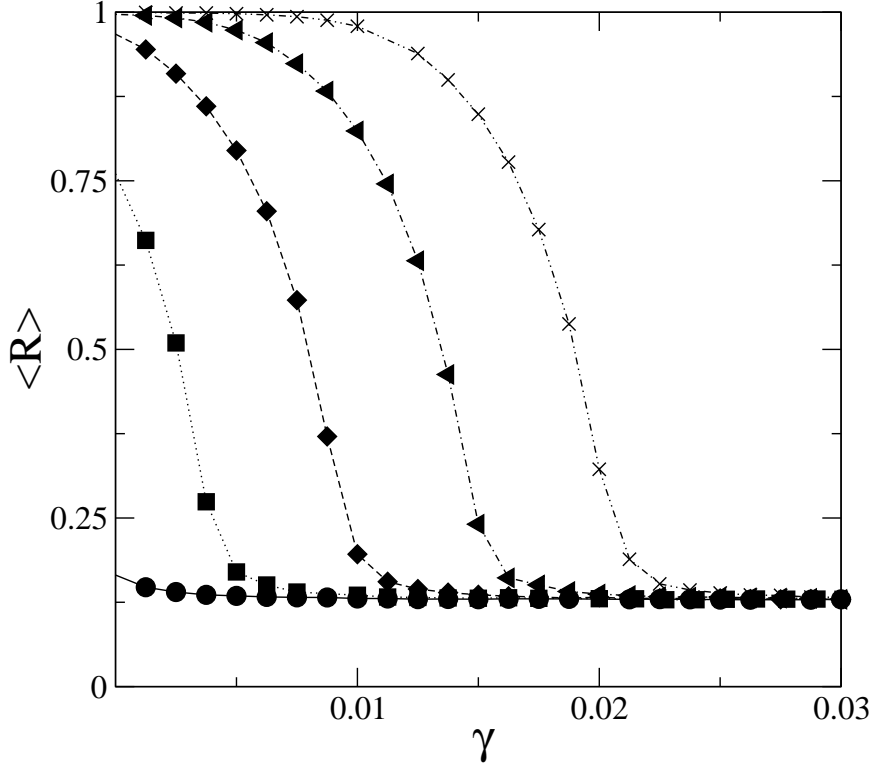}
\caption{\label{order2} Time-averaged order parameter as a function of the range parameter for $\varepsilon$ equal to $0.0125$ (circles), $0.0250$ (squares), $0.5000$ (diamonds), $0.1000$ (triangles) and $0.2000$ (crosses). We remaining parameters are the same as in Fig. \ref{campo}.}
\end{figure}

These observations can also be drawn from analyzing Fig. \ref{order2}, where the time-averaged order parameter magnitude is plotted against the range parameter $\gamma$, for different coupling strengths. If the range parameter is large enough, no coupling strengths seem to produce any synchronization at all. On the other hand, if the range parameter is zero, bursting synchronization only occurs after a critical coupling strength, what is confirmed from inspecting the curves in Fig. \ref{order2} for $\gamma$ close to zero. 

\section{Controlling bursting synchronization}

As stated in the Introduction, bursting synchronization, in the context of biological neuron assemblies, can lead to undesirable and even pathological rhythms. Hence one may think of a control strategy to suppress bursting synchronization. Such a control procedure, although in a trial-and-error basis, has been implemented in neurosurgery through deep brain stimulation \cite{tass1,tass2}. In this technique micro-electrodes are implanted in deep brain regions of a patient, like the subthalamic nucleus or {\it globus pallidus}, and a high-frequency (in the $100$-$120$ Hz range) low-amplitude signal is applied \cite{benabid}.

We will consider two strategies for controlling bursting synchronization. One of them consist on the injection of a time-periodic electric signal of low amplitude (so as not to damage the neurons) and whose ultimate goal is the suppression of bursting synchronization, leading to normal neural rhythms. The second technique is the use of a time-delayed feedback signal, with the same purpose. Each control method has its advantages in terms of the efficiency of the suppression of synchronized bursting.

The effectiveness of each control procedure on reducing or suppressing synchronization can be measured by the {\it suppression coefficient} \cite{rosenblum,rosenblum1}
\beq
\label{S}
S = \sqrt{\frac{{\mbox{\rm Var}}(X)}{{\mbox{\rm Var}}(X_f)}},
\eeq
\noindent where $X$ and $X_f$ are the values of the mean field in the absence and presence of the control, respectively. A control scheme is ideally efficient when the variance of the controlled mean field vanish, irrespectively of its value without control, corresponding thus to an infinite value of $S$. As a general rule, the larger is the value of $S$, the more efficient will be the feedback on suppressing synchronization. 

\subsection{Injection of a time-periodic signal}

We can adapt the map-based neural network model (\ref{rmx})-(\ref{rmy}) so as to model the injection of a time-periodic electric signal of low amplitude. We have implemented this procedure by adding an external time-periodic signal to just one neuron $j = S$ (the remaining neurons remaining unchanged).

Moreover, in the same way we have proceeded in the previous Section, we have chosen slightly different values for the neuron parameter $4.1 \le \alpha^{(i)} \le 4.3$. The slow timescale parameters $\sigma$ and $\beta$ have been kept fixed for all neurons, since small variations on them have caused no noticeable effect. The controlled coupled map lattice is thus 
\bea
\label{rmxc}
x_{n+1}^{(j)} & = & \frac{\alpha^{(j)}}{1 + {(x_n^{(j)})}^2} + y_n^{(j)} + \varepsilon C \sum_{\ell=1}^{N'} e^{-\gamma\Delta\ell} \left\lbrack x_n^{(j-\ell)} + x_n^{(j+\ell)} \right\rbrack + d \sin(\omega n) \delta_{j,S}, \\ 
\label{rmyc}
y_{n+1}^{(j)} & = & y_n^{(j)} - \sigma x_n^{(j)} - \beta,
\eea
\noindent where $d$ and $\omega$ are the external signal amplitude and frequency, respectively, and $\delta_{j,S}$ is the Kr\"onecker delta. The neuron $j=S$, on which the control signal is applied, can be randomly chosen. The coupled equations are also straightforwardly modified on considering situations where more than one neuron is acted upon, with a signal with the same amplitude and frequency.

We remark that the time-periodic signal is applied only at the variable representing the fast ($x$) neuron dynamics, since the slow ($y$) variable modulates the chaotic activity of $x$. For this to occur even when the neurons are coupled, it is necessary that the external inputs on the slow variables be comparatively small, a condition that cannot be warranted if the coupling is made to occur also in both variables. 

\subsubsection{Suppression efficiency}

In the following we shall denote by $\varphi^{(j)}(n)$ the bursting phase of the $j$th coupled neuron, and $\Omega^{(j)} = (\varphi^{(j)}(n) - \varphi^{(j)}(0))/n$ stands for the corresponding frequency. Besides the possibility of bursting phase synchronization, as we dealt with in the previous Section, we shall also be concerned with bursting frequency synchronization, which is a slightly weaker phenomenon since it demands that only the time-rate be equal for a number of neurons, even when their phases may differ. We also consider frequency synchronization, in the context of the present model, as an undesirable feature, and we use it so as to investigate the controllability of the time-periodic signal with a given ``external'' frequency $\omega$ \cite{rosenblum1}. 

\begin{figure}
\includegraphics[width=0.6\textwidth,clip]{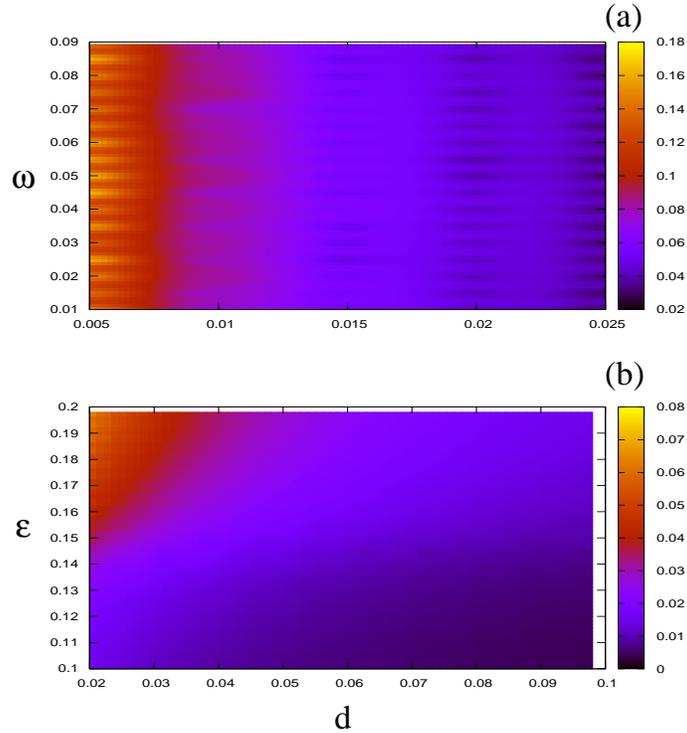}
\caption{\label{sinalnew} (color online) Suppression coefficient as a function of (a) amplitude and frequency and (b) coupling constant and amplitude, for time-periodic control signal of a network of $N = 111$ Rulkov neurons with $4.1 \le \alpha^{(j)} \le 4.3$, $\sigma = \beta = 0.001$, $\gamma = 0.005$. In (a) we used $\varepsilon=0.1$ and, in (b) we used $\omega=0.1$. The driving signal is applied to the site $S = 1$.}
\end{figure}

In Fig. \ref{sinalnew}(a) we plot (in a colorscale) the suppression coefficient given by Eq. (\ref{S}) as a function of the amplitude $d$ and frequency $\omega$ of a time-periodic control signal applied to the site $S = 1$ of a network of $N = 111$ Rulkov neurons with $4.1 \le \alpha^{(j)} \le 4.3$, $\sigma = \beta = 0.001$, $\varepsilon = 0.1$, and $\gamma=0.005$ (near zero, hence in the most favorable case since the coupling is almost global). The coupled Rulkov map lattice presents a robust synchronization of bursting, since there is little effect of the external signal on suppressing synchronized bursting. 

\begin{figure}
\includegraphics[width=0.9\textwidth,clip]{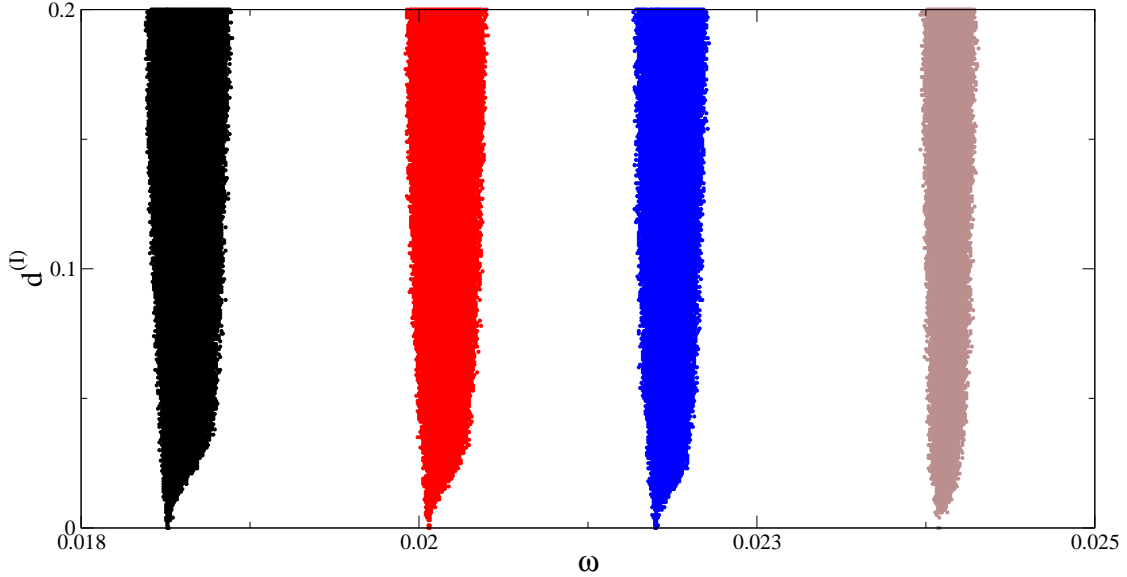}
\caption{\label{arnold} (color online) Frequency-locking ``Arnold tongues'' for bursting frequency in the external signal parameter plane of amplitude {\it vs.} frequency, for $\gamma = 0.015$ (black); $0.025$ (red); $0.035$ (blue), and $0.05$ (brown). The remaining parameters are the same as in the previous figure.}
\end{figure}

It is possible to understand this robustness of bursting by regarding the synchronized coupled map lattice as a single oscillator with many degrees of freedom being perturbed by a time-periodic forcing. Hence we expect to see a mode-locking type of phenomenon, with many ``Arnold-like'' tongues representing values of amplitude for which the system locks in the frequency of the external signal. In Fig. \ref{arnold} we show some Arnold tongues corresponding to four different values of the $\gamma$ parameter (represented by different colors). Since these tongues are of finite width, small changes in the signal amplitude cannot drive the system out of these tongues (unless we are close enough to the tongue border).
 
A comparison of interest consists on keeping the signal frequency constant and plotting the suppression efficiency as a function of the signal amplitude $d$ and the coupling strength $\varepsilon$, for different values of the $\gamma$ parameter, as before [Fig. \ref{sinalnew}(b)]. Increasing the signal amplitude does not lead to an appreciable enhancement of the efficiency on suppressing bursting synchronization. Although not shown in Fig. \ref{sinalnew}(b), we did numerical simulations for other values of $\gamma$, confirming that the synchronized behavior is rather insensitive to changes in the signal amplitude.

\subsubsection{Bursting frequency locking}

Many of the features we have seen for a time-periodic signal can be explained by interpreting its action on the network as a frequency-locking phenomenon. Hence we shall describe some aspects of this influence that are similar to the mode-locking phenomena usually described in nonlinear oscillations perturbed by harmonic forcing \cite{ottbook}.

\begin{figure}
\includegraphics[width=0.8\textwidth,clip]{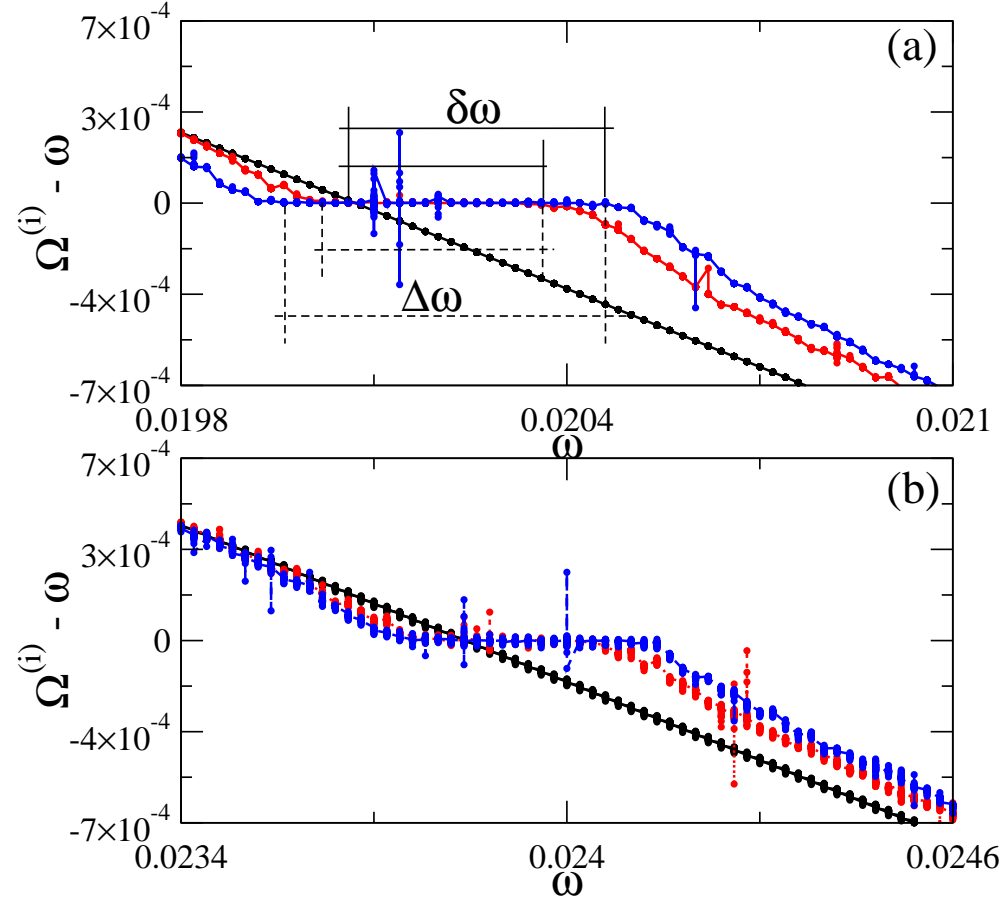}
\caption{\label{misma} (color online) Frequency mismatch of bursting neurons {\it versus} the external driving frequency for a lattice with $N = 51$ neurons, $4.1 \le \alpha^{(j)} \le 4.3$, $\sigma = \beta = 0.001$, $\varepsilon = 0.1$, and (a) $\gamma = 0.025$; (b) $\gamma = 0.050$. In both cases, the driving signal applied at the site $S = 1$ with amplitude $d = 0$ (black points), $0.05$ (red points), and $0.15$ (blue points). These points are plotted for {\it all} network sites.}
\end{figure}

We start by considering values of the coupling strength $\varepsilon$ for which the unperturbed lattice ($d=0$) exhibits bursting synchronization, with frequencies $\Omega^{(j)}$ distributed in a very narrow interval due to the distinct parameter values of each neuron. When the time-periodic signal is applied, however, the bursting frequencies $\Omega^{(j)}$ exhibit a locking with the external frequency $\omega$, what is represented in Fig. \ref{misma} as a horizontal plateau for the frequency mismatch $\Omega^{(j)} - \omega$, which is plotted against $\omega$ for {\it all} neurons belonging to the network. The concentration of these points reflect the existence of bursting synchronization, such that the latter is more suppressed as the thicker is the corresponding curve.

\begin{figure}
\includegraphics[width=0.4\textwidth,clip]{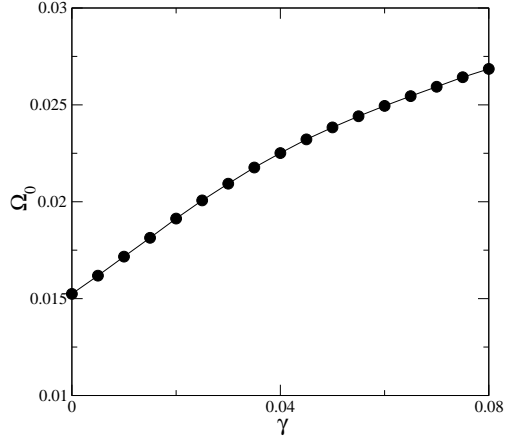}
\caption{\label{omegazero} Zero-amplitude frequency (from where Arnold tongues arise in the parameter plane) as a function of the $\gamma$ parameter for the same parameters as in the previous figures.}
\end{figure}

In this sense the ability of suppress bursting synchronization is better for higher values of $\gamma$ [Fig. \ref{misma}(b)] than for lower ones [Fig. \ref{misma}(a)], since the curves in (b) are thicker than the curves in (a). The horizontal plateaus representing frequency locking increase in both cases with the signal amplitude $d$, which is a general property of mode locking. On inspecting Fig. \ref{arnold} we see that, for different values of $\gamma$ parameter, the tongues arise from different values of the external signal frequency $\omega$. This ``zero-amplitude'' frequency $\Omega_0$ increases with the parameter $\gamma$, i.e. with couplings more and more of a locally than a globally nature, as shown by Fig. \ref{omegazero}. 

However these tongues are different from those observed in simpler systems like the sine-circle map, for example: their widths increase monotonically from both sides but with different inclinations. Moreover, after some value of $d$, the increase seems to saturate where the inclination was higher, whereas it continues to increase at the same rate where the inclination was lower. A partial explanation of this effect is that the frequency-locking interval is different with respect to the point where $d = 0$. 

On inspecting again Fig. \ref{misma} we see that the zero-amplitude frequency $\Omega_0$ of each interval is the intersection of the line black points (for which $d = 0$) and the horizontal line (where $\Omega^{(i)} = \omega$). The interval is thus wider at the righthandside of $\omega_0$ than at its lefthandside. Hence it is useful to define not only the total interval width $\Delta \omega$ but also its lefthand length $\delta\omega$. If the tongues were nearly symmetrical, we would have $\delta\omega \approx \Delta\Omega/2$. 

\begin{figure}
\includegraphics[width=0.7\textwidth,clip]{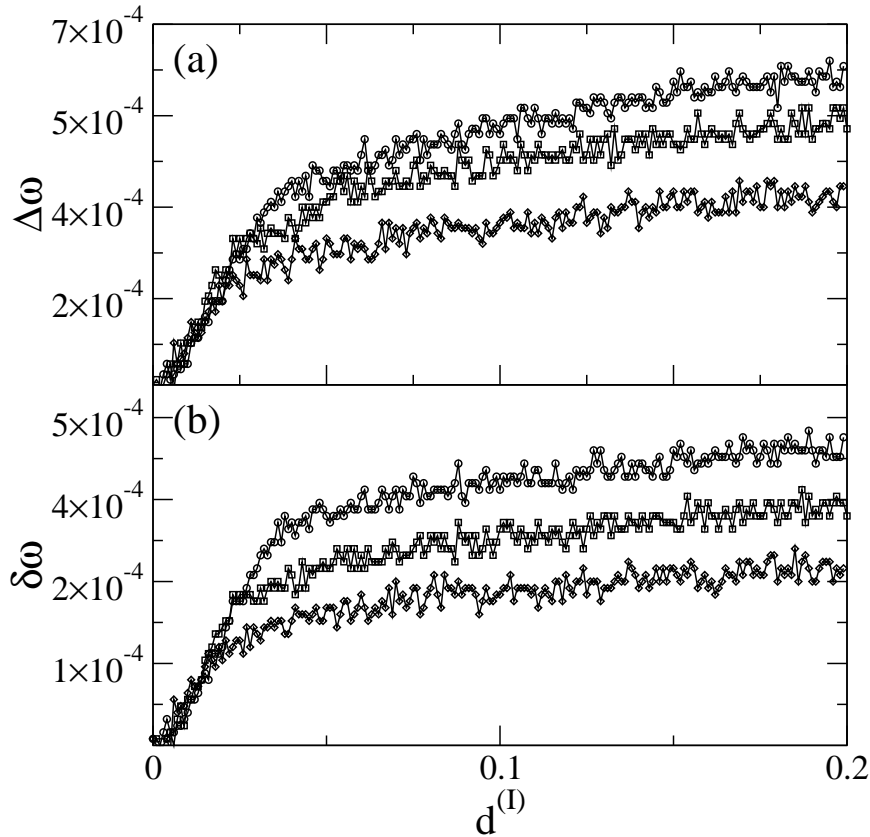}
\caption{\label{widths} Frequency locking tongue widths as a function of the signal amplitude for (a) total width $\Delta\omega$ and (b) righthand width $\delta\omega$ (with respect to the zero-amplitude frequency $\Omega_0$). In both panels we have $\gamma = 0.0125$ (circles); $0.035$ (squares); and $0.050$ (diamonds). The remaining parameters are the same as in the previous figures.}
\end{figure}

In Fig. \ref{widths} we plot both widths as a function of the signal amplitudes for different values of $\gamma$. The overall behavior of them is similar, but with some noteworthy differences: for very small signal amplitudes the tongue widths are nearly independent of the $\gamma$ value, signaling that the influence here is more from the individual dynamics than from the coupling itself. The latter manifests only after some threshold, and the widths are generally smaller for higher $\gamma$-values than for smaller ones. 

The asymmetric character of the tongues also depends on the coupling parameter $\gamma$. Let us take a fixed $\gamma$, such as $0.0125$ (represented as circles in Fig. \ref{widths}). As the signal amplitude increases both the total width $\Delta\omega$ [Fig. \ref{widths}(a)] as the righthand width $\delta\omega$ [Fig. \ref{widths}(b)] increase very mildly with $d$. However, while the former width takes on values around $5.5 \times 10^{-4}$, for large $d$, the latter is restricted to $\sim 4.5 \times 10^{-4}$, which is nonetheless an asymmetric situation. On the other hand, if we consider a larger $\gamma$, such as $0.05$ (diamonds in Fig. \ref{widths}), the total width $\Delta\omega$ is of the order of $4 \times 10^{-4}$ [Fig. \ref{widths}(a)], whereas the righthand width $\delta\omega$ is nearly half of this value [Fig. \ref{widths}(b)]. Hence for larger $\gamma$ we have more symmetrical Arnold tongues than for smaller $\gamma$, suggesting that the asymmetry of the Arnold tongues is chiefly a coupling effect, at least as the effective range is concerned. 

A qualitative explanation for the asymmetry of the mode-locking tongues has been provided by Ivanchenko et al. \cite{ivanchenko04}: an imposed positive signal precipitates a burst into a quiescent regime, and delays it when the signal is negative. If the driving frequency is higher than that of the mutually synchronized network the periodic signal will fasten the oscillations of the neurons. On the other hand, when the driven neuron starts a burst, there is an abrupt change in the mean field perceived by all neurons in the lattice, pushing them to a quiescent state. As a result, higher frequencies would give better synchronization effects.

\subsection{Delayed feedback control}

The suppression of bursting synchronization through a time-delayed feedback signal has been proposed by Rosenblum and Pikowsky \cite{rosenblum,rosenblum1}. The lattice coupling is represented by a term $\varepsilon X_n$, which includes the mean field $X_n$ given by Eq. (\ref{meanfield}). According to Ref. \cite{rosenblum} we consider two procedures of feedback control, with respect to their dependences on the mean field, described by
\begin{eqnarray}
\label{cmlpikox}
x_{n+1}^{(i)} & = & \frac{\alpha^{(i)}}{1 + {\left( x_n^{(i)} \right)}^2} + y_n^{(i)} + \varepsilon X_n + \varepsilon_f X_{n-\tau} - \varepsilon' X_n, \\
\label{cmlpikoy}
y_{n+1}^{(i)} & = & y_n^{(i)} - \sigma x_n^{(i)} - \beta, \quad (i = 1, 2, \ldots N)
\end{eqnarray}
\noindent where 
\beq
\label{contrdel}
X_n = C \sum_{\ell=1}^{N'} e^{-\gamma\Delta\ell} \left\lbrack x_n^{(j-\ell)} + x_n^{(j+\ell)} \right\rbrack 
\eeq 
\noindent and such that we have: (i) {\it direct feedback}, which takes into account the current mean field and its value $\tau$ iterations before, $X_{n - \tau}$, with coupling intensity $\varepsilon_f$, in such a way that $\varepsilon'= 0$; (ii) {\it differential feedback}, for which the controlling term is the difference between the current and the time-delayed mean fields, with $\varepsilon'= \varepsilon_f$. 

\subsubsection{Differential feedback}

\begin{figure}
\includegraphics[width=0.5\textwidth,clip]{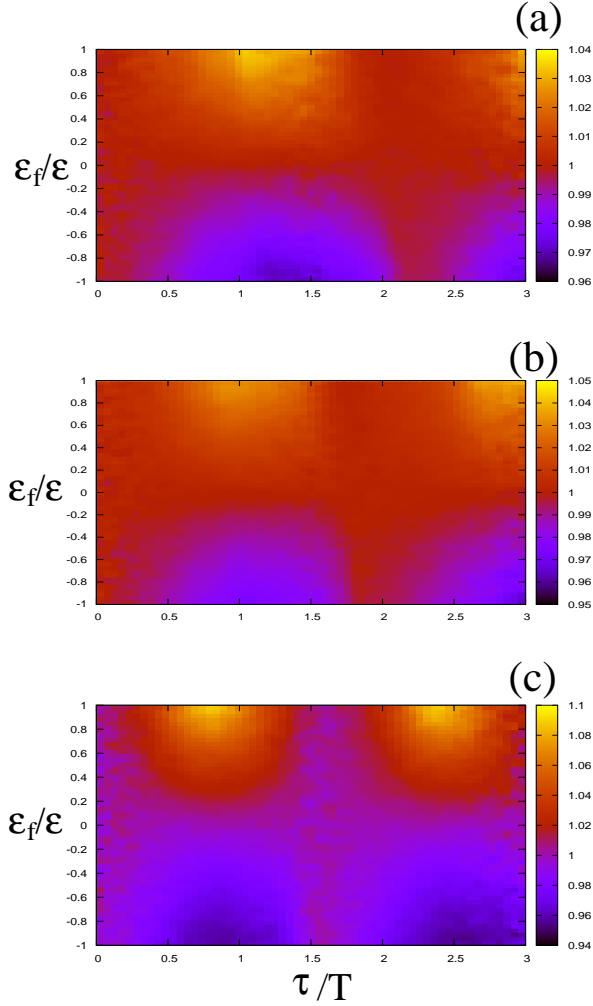}
\caption{\label{supressaodiferencial} (color online) Suppression coefficient as a function of the normalized control parameters (strength and time delay) for differential feedback control of a network of $N = 111$ Rulkov neurons with $\tau = 200$, $\varepsilon = 0.1$ and (a) $\gamma = 0.005$; (b) $0.0125$; (c) $0.025$. The remaining parameters are the same as in the previous figures.}
\end{figure}

An example of differential feedback control to suppress bursting synchronization is given in Fig. \ref{supressaodiferencial}, where we plot the suppression coefficient as a function of the control parameters $\varepsilon_f$ and $\tau$. The latter are normalized according to the coupling strength $\varepsilon=0.1$ and the mean bursting period $T=200$, respectively.

In contrast with the case of a time-periodic signal, now we observe the formation of suppression domains, or regions with large (or small) values of $S$, in the control parameter plane. As a general trend (i.e., for all $\gamma$ values considered), we observe that the domains of high (low) suppression occur for positive (negative) values of the control strength, these values being symmetric for a given value of the time delay $\tau$. 

The domains are roughly centered at values of $\tau$ which are integer multiples of the bursting period $T$. Moreover, the suppression domains (both high and low) are periodic in the time delay. Both features have been already observed by Rosenblum and Pikowsky in networks with global coupling \cite{rosenblum,rosenblum1}. In our case, we observe an influence of the parameter $\gamma$: as it increases (meaning that our coupling becomes more local) the suppression domains become centered in progressively smaller values of $\tau$ and the period in $\tau$ also decreases. On the other hand, the maximum efficiency seems not to depend on the $\gamma$, since it varies between $1.04$ [Fig. \ref{supressaodiferencial}(a)] to $1.10$ [Fig. \ref{supressaodiferencial}(c)], a mere $\sim 5\%$ of increase.

\subsubsection{Direct feedback}

\begin{figure}
\includegraphics[width=0.5\textwidth,clip]{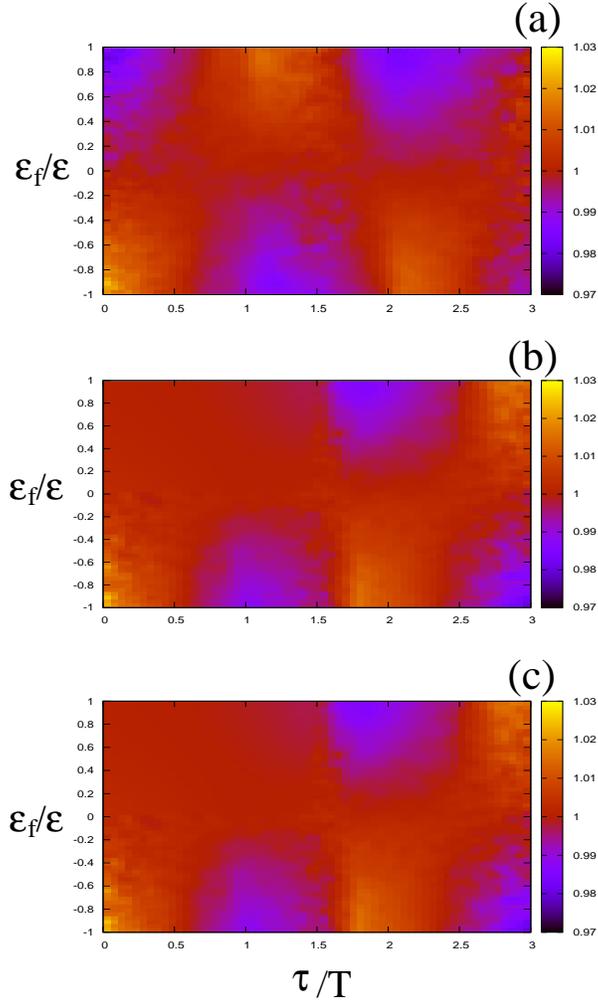}
\caption{\label{supressaodireta} (color online) Suppression coefficient as a function of the normalized control parameters (strength and time delay) for direct feedback control of a network of Rulkov neurons with (a) $\gamma = 0.005$; (b) $0.0125$; (c) $0.025$. The remaining parameters are the same as in the previous figure.}
\end{figure}

Now we consider in Fig. \ref{supressaodireta} an example of direct feedback control, with $S$ plotted against normalized $\varepsilon_f$ and $\tau$, as before. We still observe suppression domains, but with a marked contrast with the differential feedback case. Firstly the domains of low suppression are clearly more often observed than those of high suppression, showing a striking asymmetry. The domains of low suppression present three noteworthy features: (i) they occur both for positive and negative values of the control strength; (ii) they are periodic in $\tau$; (iii) they are roughly centered at multiple integers of $\tau/T$; and (iv) they are dimerized, i.e. a domain of low suppression for positive $\varepsilon_f$ is followed by another one for negative $\varepsilon_f$, after {\it circa} one period $T$. 

The location of the low efficiency domains seems not to be noticeably affected by $\gamma$, but we do observe such a domain for very small $\tau$ only in the case of globally coupled lattice (smaller $\gamma$). A curious observation is that there is a tiny domain of very high efficiency (higher than any other domain in the differential control) for negative $\varepsilon_f$, this feature observed for all $\gamma$ values considered. 

This does not mean, however, that the direct feedback is worse than the differential feedback, as far as its suppression efficiency is concerned. Comparing Figs. \ref{supressaodiferencial} and \ref{supressaodireta} we observe that the regions of moderate efficiency of the latter (between the low efficiency domains) have values of $S$ comparable with the domains of high efficiency of the differential control. However, as a whole there are more regions of high efficiency in the direct control than in differential one, the difference being that the high efficiency regions are spread out over the parameter plane in the direct feedback case. 

\section{Conclusions} 

In many biological contexts the interaction between dynamically active cells is mediated by a chemical diffusing through the inter-cell medium. This is the case, for example, of neuron synaptical connections, which are accomplished through neurotransmitters. There are two timescales involved in this kind of situations, since the timescale of the cell oscillation is generally different from the timescale related to the diffusion process. Hence a complete treatment of this kind of coupling would demand the simultaneous solution of the diffusion equation and the system of coupled oscillator equations. 

However, if the timescale related to the diffusion is much smaller than the characteristic oscillator periods we can make an adiabatic approximation and write down a closed-form expression for the nonlinearly coupled oscillator chain. For the one-dimensional case it results in a coupling term whose strength decreases with the lattice distance in an exponential fashion. An advantage of this procedure is that one can describe the commonly used global (all-to-all) and local (nearest-neighbor) cases as limiting forms of the nonlocal coupling, by varying the diffusion length that characterizes the exponential decay of interaction.

In this work we explored some of these features by using a simplified model for neuronal network consisting of map-based units displaying spiking-and-bursting behavior. Although the spiking (fast) dynamics is chaotic, the bursting (slow) dynamics presents coherent features due to the neuron interaction, one of them being their capability of synchronize the beginning of the bursting regime. 

The transition to bursting synchronization was found to depend on the nonlocal coupling features. In particular, as the inverse coupling length $\gamma$ is decreased it becomes more difficult to obtain synchronization, for a fixed coupling strength. This is compatible with the observation that locally coupled neurons are less amenable to exhibit such collective effects like bursting synchronization than globally coupled ones.

This is particularly interesting from the point of view of potential disorders (like abnormal rhythms associated with epilepsy and Parkinson disease) related to them. In order to control those undesirable bursting synchronized rhythms we have used two strategies for suppressing bursting synchronization: a time-periodic signal and a delayed feedback one. In the former procedure a time-periodic harmonic and low-intensity signal injected on a randomly chosen neuron. This external signal is capable to change the bursting frequencies of interacting neurons so as to drive them out of the synchronized behavior. The inclusion of a time-periodic signal causes the synchronized neurons to lock into an Arnold-like tongue with a well-defined width. 

The common bursting frequency was found to increase with the inverse coupling length. Moreover, the frequency locking tongues were found to be asymmetric with respect to this common locking frequency. As a consequence, if one injects an external signal with frequency higher than the upper limit of the frequency-locking tongue we can obtain desynchronization, as required. Both the amplitude and the asymmetrical features of the tongues were found to depend on the nonlocal character of the interaction between neurons. 

In terms of the efficiency of the bursting synchronization suppression, we did not find an appreciable effect of an external time-periodic signal of constant amplitude and frequency, even after wide changes in these parameters. Accordingly, we sought for a non-constant scheme of control, preferably with an amplitude tailored to suit the actual needs of suppressing synchronized behavior. This was accomplished by the injection of a time-delayed feedback signal. In contrast with the previous control scheme, for this feedback signal we observed formation of suppression domains for time delays roughly centered at multiples of the bursting period, with a well-defined periodicity. 

Two feedback control types were used: a differential scheme, with a relative symmetry between domains of low and high suppression, and a direct scheme, with an apparent bias towards low suppression domains. On the other hand, we found that the direct scheme seems superior to both a time-periodic signal and a differential feedback signal in the sense that the suppression efficiency is higher over the control parameter plane, even though we do not observe many domains of control for this case. Hence, from a ``practical'' point of view, a direct feedback delayed signal would yield better results in terms of the suppression of undesired (and pathological) synchronized rhythms.

The extension of the present formalism to higher-dimensional lattices is quite straightforward, but the coupling kernel is somewhat different and the summation through all neighborhoods may become a computationally difficult task. However, the general features of such a kind of coupling are already present in the simpler one-dimensional case. 

\section*{Acknowledgments}

This work was made possible with help of CNPq, CAPES, and Funda\cao Arauc\'aria (Brazilian Government Agencies).

\end{document}